\documentclass[letterpaper, conference]{IEEEtran}

\usepackage{graphicx} %
\usepackage{subcaption}
\usepackage{mathptmx} %
\usepackage{amsmath,bm} %
\usepackage{amssymb}  %
\usepackage{textcomp}
\usepackage{array,multirow}

\begin{document}

\title{ Learning Sensor Placement from Demonstration \\
for UAV networks \\
}

\author{\IEEEauthorblockN{Assia Benbihi}
\IEEEauthorblockA{\textit{UMI2958 GeorgiaTech-CNRS }\\
\textit{CentraleSup{\'e}lec, Universit{\'e} Paris-Saclay}\\
abenbihi@georgiatech-metz.fr}
\and
\IEEEauthorblockN{Matthieu Geist}
\IEEEauthorblockA{\textit{Universit{\'e} de Lorraine } \\
\textit{and CNRS,LIEC, UMR 7360}\\
Metz, France 
}
\and
\IEEEauthorblockN{C{\'e}dric Pradalier}
\IEEEauthorblockA{\textit{UMI 2958, GeorgiaTech-CNRS}\\
\textit{GeorgiaTech Lorraine}\\
Metz, France
}
}
\maketitle

\begin{abstract}

  This work demonstrates how to leverage previous network expert demonstrations
  of UAV deployment to automate the drones placement in civil applications.
  Optimal UAV placement is an NP-complete problem:
  it requires a closed-form utility function that defines the environment and
  the UAV constraints, it is not unique and must be defined for each new UAV
  mission. This complex and time-consuming process hinders the development of
  UAV-networks in civil applications. We propose a method that leverages
  previous network expert solutions of UAV-network deployment to learn
  the expert's untold utility function form demonstrations only. This is especially
  interesting as it may be difficult for the inspection expert to explicit his
  expertise into such a function as it is too complex. Once learned, our
  model generates a utility function which maxima match expert UAV
  locations. We test this method on a Wi-Fi UAV network application inside a
  crowd simulator and reach similar quality-of-service as the expert. We show
  that our method is not limited to this UAV application and can be extended to
  other missions such as building monitoring.

\end{abstract}


\section{Introduction}
Following the Google projects Loon, Skybender and the Facebook project Aquila,
a small scale of UAV networks can be deployed for coverage where UAV are used
as communication relays. The survey \cite{hayat2016survey} reports diverse
civil applications such as communication networks in regions with few resources
or areas that suffered natural disasters, search and rescue missions or
long-term environment monitoring. Several challenges are discussed there among
which communication protocols, robot collaboration and autonomous navigation.
In this paper, we focus on the UAV placement strategy which differs for each
civil application, each UAV mission and each network technology. This
dependency hinders the definition of a general deployment strategy and
systematically requires a network expert to specify the UAV locations. We
leverage the expert's effort to build an automatic UAV deployment model from
expert demonstrations.

\begin{figure}[htb]
  \centering
      \includegraphics[width=0.7\linewidth]{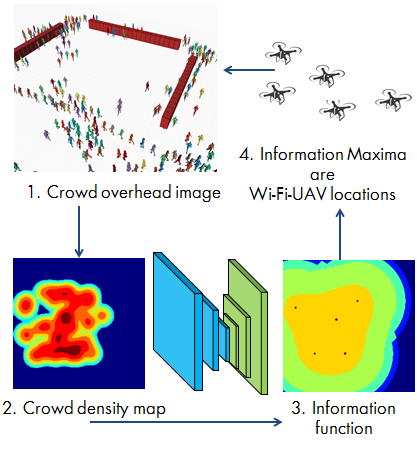}
      \caption{The crowd map is converted to a density map 
      fed to our model to generate a 2D utility function. Optimal
      sensor locations are the local maxima of this function.}
      \label{fig:method}
\end{figure}

Drone placement is a specific case of optimal sensor
placement under constraints which is an NP-complete problem. Existing
approaches frame it either as a geometric or a statistical optimization. In the
first case, the task becomes an instance of the Art-Gallery-Problem (AGP)
\cite{o1987art} where sensors have a fixed sensing radius and must be placed to
maximize the surface covered~\cite{gonzalez2001randomized}. The statistical
approach aims at modeling the mission with Gaussian processes that the expert
must tune. He then uses the model to define an optimization criteria over the
sensor locations. Among them are the location entropy
\cite{shewry1987maximum} or the mutual information between a location and
non-sensed locations \cite{krause2008near}. For both approaches, an important
part of the resolution efforts lies in defining the utility function that
embeds the infrastructure model, the coverage requirements and the physical
world constraints. Not only this definition must be reiterated for each new
mission, but it is also complex for the network expert to embed his expertise
into it. Instead, we endeavor to learn this utility function
from expert demonstrations of sensor placements on previous missions.

\begin{figure*}[htb]
  \centering
      \includegraphics[width=0.6\linewidth]{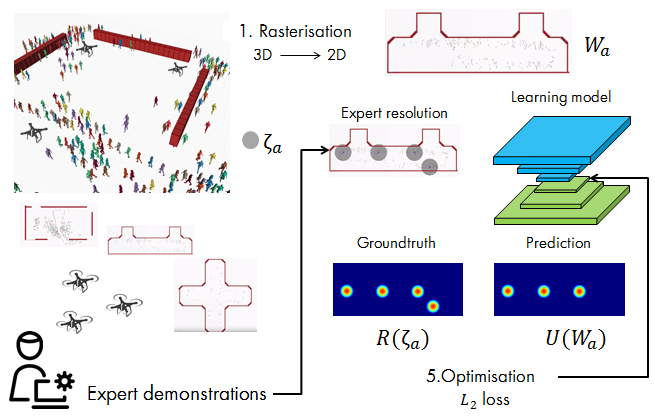}
      \caption{The CNN is trained by comparing its utility function
      \textbf{maxima} with the sensor locations of the expert.}
      \label{fig:training}
\end{figure*}

To do so, we define a learning model that takes an overhead image of an
infrastructure and outputs a 2D utility function of which maxima are optimal
sensor placements according to the expert. Figure \ref{fig:method} illustrates
our approach on a simulated crowd for which Wi-Fi UAVs provide an internet
connection. We choose a Convolutional Neural Network (CNN) to represent our
utility function generator as it can process raw visual information contrary to
other models that require hand-crafted features. We rely on Learning from
Demonstration (LfD) \cite{argall2009survey} to train the CNN and Figure
\ref{fig:training} illustrates the optimization: we compare the local maxima of
the CNN utility function together with the expert locations and penalize the
network proportionally to how far they are from each other. The CNN constrains
us to do so in a differentiable way so we introduce an original regression loss.
We simplify this learning into a simple regression problem with an astute
projection of the expert locations into a 2D Gaussians. The penalty
can then be estimated with a simple differentiable Mean-Square-Error (MSE) loss
between this map and the CNN output. This conversion is inspired by the expert
demonstration representation in visual attention model \cite{borji2013state} 
(section II-B).

Our method is tested in a crowd simulator in which we add a simple wireless
communication model. We choose two coverage strategies and implement an expert
oracle to generate sensor location examples.  Our approach reaches the same
Quality of Service (QoS) as the oracles on both strategies. We also show that
our method is not specific to the input data representation and solve two
building monitoring problems based on the AGP and the Fortress Problem (FP).

Our contribution lies in the optimization method to learn a utility
function only from demonstrations to solve sensor placement problems
under constraints. This proof of concept targets the definition of the data
representation and the learning optimisation. It is part of the global effort
aiming at relieving the expert from modeling the coverage mission and
explicitly defining a complex utility function.

\section{Problem Formulation}
This section frames the sensor placement optimization in deep learning
from demonstration. Given a 2D representation of the UAV mission, our model generates 
a utility function of which maxima are the UAV locations. These locations are
chosen so as to mimic the network expert placements. To do so, the model
is trained on examples of optimal sensor placements that the network expert
provides.

\subsection{The Learning Problem Definition}
Let $a$ be a UAV mission rasterised into a 2D map 
$W_a\in\mathbb{R}^{h \times w}$ and $\zeta_a \in \mathbb{R}^{N_a \times 2}$ 
be the $N_a$ expert locations on $W_a$.
We endeavor to learn a utility function generator $U$ on mission instances $W_a$ 
so that the maxima of the utility function $U(W_a)$ are $\zeta_a$:
\begin{equation}
  \begin{split}
  \begin{array}{lrlrl}
    U : &  \mathbb{R}^{h\times w} & \rightarrow & \mathbb{R}^2 & \rightarrow \mathbb{R}\\
        &   W_a                   & \rightarrow & U(W_a) : (i,j) & \rightarrow   U(W_a)(i,j)\\
  \end{array} \\
    \textrm{such that argmax}_{\zeta \in \mathbb{R}^{N \times 2}, N \in \mathbb{N}} U(W_a)(.) = \zeta_a
\end{split}
    \label{eq:problem}
\end{equation}

We choose to represent $U$ with a CNN as it is a non-linear function with a
powerful representation space and can process raw images contrary to other
models that require hand-crafted features. One constraint of the CNN is that
its loss function must be differentiable. An intuitive loss is to measure how
well the expert and learned locations $\zeta_a$ and $\hat{\zeta_a}$ align. To
do so, we first need to compute $\hat{\zeta_a}$ i.e. $U(W_a)$'s maxima. Then we
solve the assignment problem between $\hat{\zeta_a}$ and $\zeta_a$ to assign
each location in $\hat{\zeta_a}$ to the nearest expert location in $\zeta_a$.
Finally, the loss is the Euclidean loss between the pairs assigned in the
previous step. However, neither the argmax operation nor the assignment
optimization are differentiable operations. Another difficulty comes from the
varying number of $U(W_a)$ maxima which is common to any learning problem with
a list for output. Also, a trivial solution is to put sensor everywhere, which
we want to avoid. To answer all these challenges, we define an alternative
differentiable loss that still follows this intuitive behavior and casts the
learning problem as a simple regression.

\subsection{The Loss Definition}

We define a differentiable loss $\mathcal{L}$ that penalizes the CNN with
respect to how far matching locations in $\hat{\zeta_a}$ and $\zeta_a$ are. It
must meet the following requirements:
\begin{itemize}
  \item $\mathcal{L}$ must be differentiable.
  \item $\mathcal{L}$ must be an increasing function of the euclidean distance
    between matching location pairs in $\hat{\zeta_a} \times \zeta_a$.
  \item Modularity condition: $\mathcal{L}$'s minimization residue decreases 
    as $U(W_a)$ has more maxima.
\end{itemize}
To do so, we convert an expert demonstration $\zeta_a$ into a saliency map
$R(\zeta_a) \in \mathbb{R}^{h\times w}$: for each location in $\zeta_a$, we
draw an unnormalized Gaussian of mean $p$ and variance $\sigma$ around each
expert position $p$ (Figure \ref{fig:training}: $R(\zeta_a)$). This constrains
the CNN to also produce a saliency map $U(W_a)$ with high peaks at sensor
locations. The training loss then measures the distance from the expert
saliency map $R(\zeta_a)$ to the CNN one $U(W_a)$:
\begin{equation} \label{eq:lossDfn}
  \begin{split}
    \mathcal{L}(U(W_a)) &= \| R(\zeta_a) - U(W_a)\|_2 \\
    \textstyle \textrm{with} \quad R(\zeta_a)(i,j) &= \textstyle \sum_{p \in \zeta_a} \alpha exp\frac{-\|p - (i,j)\|^2}{2\sigma^2}
  \end{split}
\end{equation}
This loss can be interpreted as `how well $\hat{\zeta_a}$ and $\zeta_a$ align'.
The conversion of $\zeta_a$ into a saliency map is inspired by the visual
attention model data representations \cite{borji2013state}. As for the loss
definition, it is motivated by the expert resolution of the task that can be
cast as a submodular function $f$ optimization:
\begin{equation} \label{eq:expertOpt}
\textrm{max}_{\zeta \subseteq \mathbb{R}^2 } f(\zeta) \quad \textrm{subject to some constraints on} \quad \zeta
\end{equation}
The submodularity property means that for a set of positions $\zeta$ and $\zeta'$ such
that $\zeta \subseteq \zeta'$, a new position $p$: 
\begin{equation}
  f(\zeta \cup \{p\}) - f(\zeta) \geqslant f(\zeta' \cup \{p\}) - f(\zeta')
\end{equation}
Intuitively, this diminishing-return condition means that the more sensors one
places, the less additional information the new sensor provides. An example of
the submodularity condition is the following: let $\zeta = \emptyset$ and
$\zeta'$ a set of $U(W_a)$ maxima. If the model produces a new maxima $p$, the
loss diminishes more if one starts from $\zeta=\emptyset$ than when starting
from $\zeta'$. 

Regarding the tuning of the $\sigma$ parameter, the Gaussian variance $\sigma$
must be small enough so that the map is only maximal at $\zeta_a$ with no
residual local maxima. Also, the smaller the variance, the higher is the
penalty when the utility function $U(W_a)$ misses a maxima or places it far
from $\zeta_a$. However, a $\sigma$ too small creates wide areas where the
penalty signal is not informative enough for the network to learn. So $\sigma$
must also be high enough to create a non-null learning signal everywhere to
avoid that the model gets stuck between two expert sensor locations. A
solution is to use two Gaussians: a first one with a small variance and a high
amplitude which goal is to penalize misplaced $U(W_a)$ maxima. And a second
Gaussian with a higher variance and a smaller amplitude to strengthen the
learning signal and guide $U(W_a)$ maxima that are far from $\zeta_a$ towards
them. The loss now depends on four parameters $(\alpha_1, \sigma_1),
(\alpha_2, \sigma_2)$:
\begin{equation} \label{eq:lossDfn2}
  \begin{split}
    \mathcal{L}(U(W_a)) &= \| R(\zeta_a) - U(W_a)\|_2 \\
      \textstyle \textrm{with} \quad R(\zeta_a)(i,j) &= \textstyle \sum_{i\leqslant 2} \alpha_i \sum_{p \in \zeta_a} exp\frac{-\|p - (i,j)\|^2}{2\sigma_i^2}
\end{split}
\end{equation}
In practice, we set $(\alpha_1=255, \sigma_1=1)$. Then we perform a grid search
to choose $(\alpha_2, \sigma_2)$. The grid search range depends on the expert
solution distribution: in general, $(\alpha_2, \sigma_2)$ are smaller when the
mission usually requires more sensors. 
We assume that there may be several
solutions for one task but only one solution strategy. The incorporation of
multiple strategies is not addressed here. 

\subsection{Utility Function as a Convolutional Neural Network}

This section details the representation of the utility function generator $U$.
It is modeled by a convolutional neural network (CNN) with an encoder-decoder
structure. It can learn from raw visual data and avoids the design of
hand-crafted features. CNN are made of a composition of filter banks trained to
extract relevant features. The only learning signal it needs to train is the
expected output of the CNN. In this case, it is the rasterised expert
solutions $R(\zeta_a)$. 

The encoder-decoder CNN structure is divided in two parts. The encoding part
(Figure \ref{fig:training}, blue) of the network is made of multiple filter
banks interleaved with max-pooling layers that reduce the spatial dimension 
while keeping the most relevant features. The output of the encoder is the
summary of the input features at multiple scales. It is fed to a decoder
(Figure \ref{fig:training}, green) which generates a utility function out of
the learned relevant features. The decoder is also made of filter banks but
they are interleaved with upsampling layers \cite{zeiler2011adaptive} so that
the output of the decoder have the same spatial dimension as the input. In the
decoder, the filters banks are transpose convolutions \cite{dumoulin2016guide}
that computes an embedding of the relevant features into a utility function. 

The filter size and the number of layers in the encoder must be set together so
that the scope of the last encoding layers covers the entire spatial dimension
of the input. For an input image of size 64 pixels and max-pool layers every
two convolutional layers, this can be achieved with a filter size of 3 and 4
convolutional layers. We advise to use the minimum number of encoding layers
as possible since the bigger the network, the more expert demonstration you
need to train it without overfitting. Also in practice, we observe that a
bigger network may not even converge. Another good practice is to use a stack
of small kernel filters rather than one big kernel
filter~\cite{simonyan2014very}.

Once the encoding part is set, there are two possible ways to decode: either
make a heavy decoder symmetric to the encoder \cite{badrinarayanan2017segnet}
or make a light decoder of only upsampling layers, one for each pooling layer
in the encoder. The heavy decoder uses much more bank filters than the second
but may be relevant for task with highly informative inputs which need more
filters to extract all the relevant features. The light decoder proves to be
enough to learn $U(W_a)$ on our two use cases.

\begin{table}[h]
  \scriptsize
  \caption{Neural Network model.}
    \label{tab:network}
    \begin{center}
      \begin{tabular}{|c|c|c|c|c|c|c|}
        \hline
        Encoder & 2 $\times$ conv-32      & P   & 2 $\times$ conv-64    & P   & 2 $\times$ conv-128 \\
        \hline
        Decoder & 1 $\times$ conv-32$^T$  & UP  & 1 $\times$ conv-64$^T$& UP  &1 $\times$ conv-128$^T$ \\
        \hline
      \end{tabular}
    \end{center}
\end{table}

We note $N \times \textrm{conv-}C$ the composition of $N$ banks of $C$ 2D convolution
filters, and $N \times \textrm{conv-}C^T$ when we use transpose-convolutions
\cite{dumoulin2016guide}.
The max-pooling and upsampling layers \cite{zeiler2011adaptive} are designated by `P' and `U'
respectively. The encoder and decoder architectures are summarized in Table 
\ref{tab:network}.

\begin{table}[h]
  \caption{Training parameters.}
    \label{tab:trainingParam}
    \begin{center}
        \begin{tabular}{|c|c|c|}
        \hline
                          & Wi-Fi  & Monitoring \\
        \hline                     
        Train set size    & 5000   & 25000    \\
        \hline                     
        Test set size     & 800    & 1000     \\
        \hline                     
        $W_a$ image size  & 256    & 64       \\
        \hline                     
        Batch size        & 10     & 64       \\
        \hline                     
        Train epochs      & 500    & 300      \\
        \hline                     
        Training duration & 6h     & 4h       \\
        \hline
      \end{tabular}
    \end{center}
\end{table}

The networks are implemented using the TensorFlow \cite{abadi2016tensorflow}
framework on a Nvidia GT1080 GPU. The networks are trained with the Adam
optimizer \cite{kingma2014adam} and the other training parameters are summarized in Table
\ref{tab:trainingParam}. 

\section{Experiments}
This section describes the simulation and the training.
\subsection{UAV Wireless Network Simulation}

\begin{figure}[htb]
  \centering
      \includegraphics[width=0.8\linewidth]{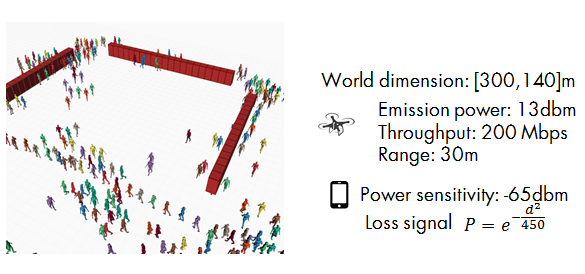}
      \caption{Pedsim crowd simulation rendered in ROS. 
      The grey circles represent the network coverage of each UAV.}
      \label{fig:uav_simu}
\end{figure}

A simple communication model is defined within the
Pedsim~\footnote{http://pedsim.silmaril.org/} crowd simulator wrapped in the
Pedsim-ROS package~\footnote{https://github.com/srl-freiburg/pedsim}. We
generate fake scenarios by defining random obstacles and waypoints for agents
and run the simulation (Figure \ref{fig:uav_simu}). The drones are equipped
with a wireless AC card with a maximum throughput of 200Mbps, a signal range of
30m, and an emission power signal of 13dbm. These are standard values for Wi-Fi
routers. The end receptor power sensitivity is set to -65dbm, which is
enough to stream videos and run VoIp calls with a smartphone.  We approximate
the free-space path loss with an unnormalized Gaussian
$e^{-\frac{d^2}{2\sigma^2}}$ after rescaling the power signal from [-65dbm,
13dbm] to [0.1, 1]. $\sigma^2$ is computed so that the signal power at $d$=30m
is 0.1 so $\sigma^2=450$. 

We define two UAV deployment strategies summarized in Table
\ref{tab:UAVpolicies}. The first one should place the UAVs to cover the maximum
number of agents without regards to the end-user throughput bandwidth. The
second one should cover the maximum number of agents under the constraint of
providing them with a minimum of 5Mbps throughput, which is more than enough to
watch Netflix with ultra HD quality. In both cases, the expert deploys a drone only if
there is a cluster with more than 30 end-users without network access. We
ignore other communication network related issues addressed
\cite{guillen2016wifi} to focus only the placement problem.%

\begin{table}[h]
  \caption{UAV deployment strategies}
    \label{tab:UAVpolicies}
    \begin{center}
        \begin{tabular}{|p{3.5cm}|p{3.5cm}|}
        \hline
        \multicolumn{1}{|c}{Strategy 1 (S1)} &\multicolumn{1}{|c|}{Strategy 2 (S2)} \\
        \hline
        \multicolumn{2}{|c|}{Maximum number of UAV: 10}  \\
        \hline
        \multicolumn{2}{|c|}{Minimum size of group worth covering: 30}  \\
        \hline
        \multicolumn{1}{|c|}{-} & bandwidth $\geq 5Mbps/\textrm{agent}$ \\
        \hline
      \end{tabular}
    \end{center}
\end{table}

\subsection{UAV Monitoring Simulation}
\begin{figure}[htb]
  \centering
      \includegraphics[width=5cm]{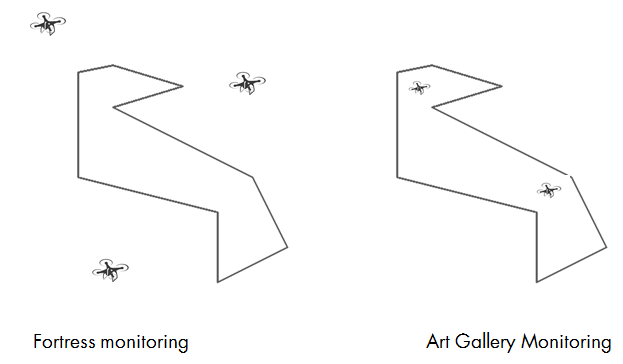}
      \caption{Left: FP, the UAV cover the exterior perimeter.
      Right: AGP, the UAV cover the interior surface.}
\label{fig:agp_simu}
\end{figure}

In addition to the communication coverage above, we show that our method is
restricted by neither the input data representation nor by the expert utility.
To this end, we test it on a buliding monitoring task with UAV networks. We
simulate the buildings with random polygons and emulate two experts with the
famous Art-Gallery-Problem (AGP) and Fortress-Problem (FP) solutions (Figure
\ref{fig:agp_simu}).
These two sensor placement problems pertain to a broad category of problems named
Next-Best-View (NBV) problems. They arise in many applications among which
autonomous robotic exploration, environment monitoring and inspection. 
In AGP, the expert places the
minimum number of UAV-guards to always monitor the interior of an art gallery
and catch potential thieves. In FP, the expert deploys the minimum
number of guards outside of a fortress to constantly watch its perimeter. More
generally, we optimize a visibility coverage function while minimizing the
number of guards. For both missions, buildings are modeled with random
polygons~\cite{cgal:dhhk-gog-17a} and the expert solutions are computed with
the numerical solver~\cite{tozoni2013practical}.  In this context, $W_a$
(eq.~\ref{eq:problem}) is a polygon drawing on a $[64 \times 64]$ pixels map. 

\subsection{Evaluation}

We introduce a generic metric to evaluate the utility function generator $U$:
the \textit{correspondence ratio}. It measures the proportion of expert
locations $\zeta_a$ \textit{matched} by the placements $\hat{\zeta_a}$ from the
learned policy. We define \textit{matching locations} as follow: we first solve
the assignment problem between $\zeta_a$ and $\hat{\zeta_a}$, and note the
resulting location pairs $\mathcal{P}$. Let $(p,\hat{p}) \in \mathcal{P}$, $p$
and $\hat{p}$ match when $\|p - \hat{p}\|_2 \leqslant \epsilon_a$, with
$\epsilon_a$ a distance threshold specific to the task dimension. We set
$\epsilon_a$ at 10\% of the mission dimension and define the correspondence
ratio $corr$ as $corr = \frac{|match|}{|\zeta_a|}$.
We control whether $U(W_a)$ generates trivial solutions with a score inspired
from the {\em precision} metric in classification, hence the score name we use. 
We note $\hat{\zeta_a}^*$ the subset of $\hat{\zeta_a}$ that match no
expert sensor locations and define the precision as
$prec = \frac{|match|}{|match| + |\hat{\zeta_a}^*|} $. 
The score is 100\% when the learned placements match exactly the expert
ones with no additional locations then it decreases as the number of $U(W_a)$
maxima increases.

Contrary to the training loss, we do not penalize $U$ if $U(W_a)$ has more
maxima $\hat{\zeta_a}$ than the expert solution $\zeta_a$. In practice, we
observe that these additional maxima are other valid solution for the task.
This metric is completed with task specific metric such as QoS. In the first
strategy of the Wi-Fi UAV simulation, we measure the QoS with the proportion of
end-users covered by the Wi-Fi network. For the second Wi-Fi UAV strategy, we
also require the end-users's throughput i.e. the amount of data received per
second, to be higher than 5Mbps. As for the monitoring problems, the `QoS` is
the ratio of buildings for which perfect coverage is reached. The expert always
generates UAV locations that reach a correspondence ratio and QoS of 100\%.

\section{Results}
The learned UAV placements $\hat{\zeta_a}$ and $\zeta_a$, the expert's ones, have
at least 75\% of correspondences for all simulations. Positions correspond if
they are closer than 10\% of the simulation dimensions. There is substantial
room for improvement regarding the exact matching of learned positions (25\%).
However, the QoS metrics suggest that although the learned locations differ
from the expert ones, they are still valid deployment strategy and offer high
QoS. Indeed, for all simulations, the QoS is higher than the correspondence
ratio which shows that when the learned utility function does not mimic
exactly the expert's one, its maxima are an alternative valid solutions to
the sensor placement problem. We study the influence of the learning parameter
$\alpha_2$ and observe that it improves the correspondence ratio up to 29\%.

\subsection{UAV Wi-Fi network}

\begin{table}[h]
  \caption{Correspondence ratio and QoS.}
    \label{tab:all_res}
    \begin{center}
      \begin{tabular}{|c|c|c|c|c|}
        \hline
        Strategy  &  Corresp. (\%)  & QoS (\%)  & Precision (\%) \\
        \hline                              
        UAV 1     & 75              & 100       & 75 \\ 
        \hline
        UAV 2     & 89              & 93        & 72    \\
        \hline                              
      \end{tabular}
    \end{center}
\end{table}

The correspondence ratio is above 75\% for both UAV deployment strategies. It
even reaches 89\% for the second one where placements are additionally
constrained with the end-user throughput. This supports the hypothesis that $U$
embeds the expert knowledge on the UAV placement and aims at imitating it as
much as possible. 

We also observe that the QoS is always higher than the correspondence ratio
which shows that even when $\hat{\zeta_a}$ and $\zeta_a$ do not match,
$\hat{\zeta_a}$ still is a valid solution. This reinforces the hypothesis that
$U$ can not only imitate the expert, but it can also provide alternative
solutions while still abiding by the expert utility it learnt.
Thus, the learned deployment still completes the mission even when it is
different from the expert's one. For the first strategy, 100\% of the clusters
with more than 30 end-users are covered even though the learned placements
match only 75\% of the expert's ones. The model trained on the second strategy
also covers all end-users clusters and misses the throughput requirement for
only 7\% of the end-users. Qualitatively, we observe that the utility
function $U(W_a)$ tends to either have maxima that match exactly the expert or
produce a higher number maxima. It is up to the non-maxima-suppression step
($argmax_{\zeta \in \mathbb{R}^{N \times 2}} U(W_a)$) to then filter
utility maxima among those proposed.

\begin{figure}[htb]
\centering
\includegraphics[width=0.5\linewidth]{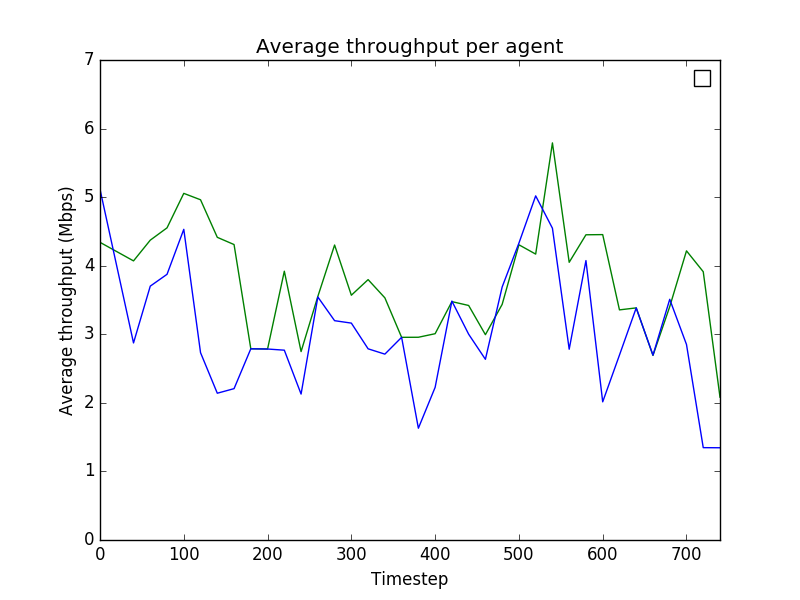}
\caption{Average end-user throughput over time during a dynamic sequence.
  Green: Expert placements $\zeta_a$. Blue: Learned placements $\hat{\zeta_a}$.
  $\hat{\zeta_a}$ and $\zeta_a$ lead to the same throughput pattern.}
\label{fig:uavThroughput}
\end{figure}

Figure \ref{fig:uavThroughput} shows the average throughput per agent over time
during one dynamic scene for the expert (green) and the learned placement
(blue). We observe that the learned placements produce a throughput pattern
that follows the one generated with expert sensor locations. This suggests that
$U$ not only embeds the expert utility function a time $t$ but also its
dynamics. 

\subsection{UAV Monitoring Network}
\begin{table}[h]
  \caption{Correspondence ratio and QoS.}
    \label{tab:all_res}
    \begin{center}
      \begin{tabular}{|c|c|c|c|}
        \hline
        Strategy  &  Corresp. (\%)  & QoS (\%)  & Precision (\%) \\
        \hline                              
        AGP       & 91              & 97        & 82 \\ 
        \hline
        FP        & 86              & 90        & 78    \\
        \hline                              
      \end{tabular}
    \end{center}
\end{table}

The model learnt on the monitoring demonstrations comforts the previous
analysis. It reaches a correspondence ratio above 86\% on both missions (Table
\ref{tab:all_res}). This shows that the learning method can generalise to
different data representations. It also suggests that $U$ can embed high-level
concepts, such as coverage, even when the data holds only a few visual
features. We compute the `QoS' coverage as the ratio of buildings for which
perfect coverage is reached. As with the previous simulation, the QoS is higher
than the correspondence ratio. This strengthens the observation that even when
$U(W_a)$ maxima do not match the expert solution $\zeta_a$, they still provide
another valid solution.

\subsection{Tuning the Learning Signal: $\alpha_2$ setting}

\begin{figure}[thb]
  \centering
  \makebox{
    \parbox{0.45\linewidth}{
      \hbox{ \includegraphics[width=\linewidth]{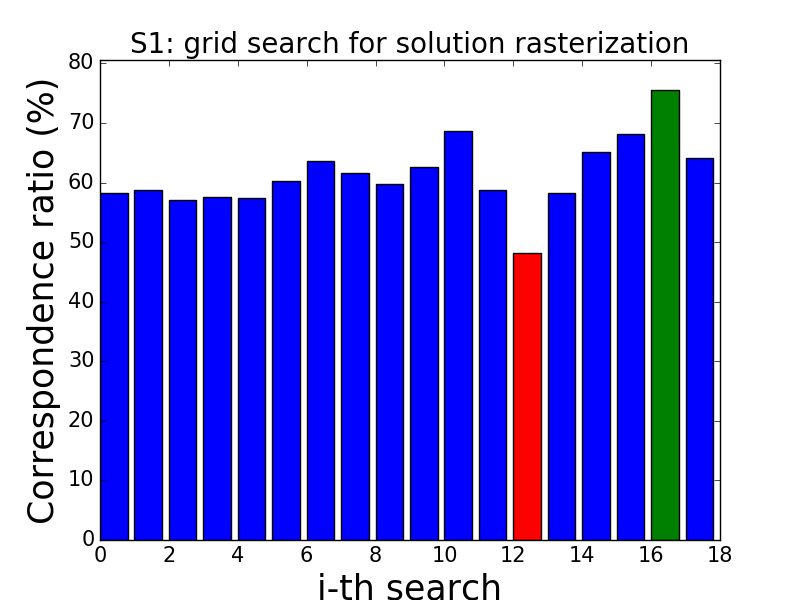} }
      \hbox{ \scriptsize \hspace{0.7cm} UAV 1: $\alpha_2=50, \sigma_2=50$}
      \hbox{ \scriptsize \hspace{1.5cm} $\alpha_2$-rise=21\% }
      \hbox{ \includegraphics[width=\linewidth]{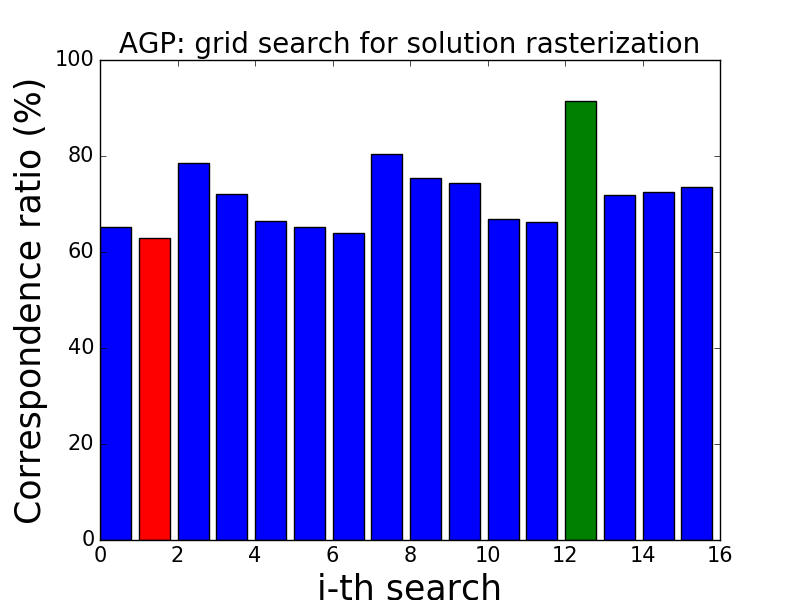} }
      \hbox{ \scriptsize \hspace{0.7cm} AGP: $\alpha_2=40, \sigma_2=40$}
      \hbox{ \scriptsize \hspace{1.5cm} $\alpha_2$-rise=29\% }
    }
    \parbox{0.45\linewidth}{
      \hbox{ \includegraphics[width=\linewidth]{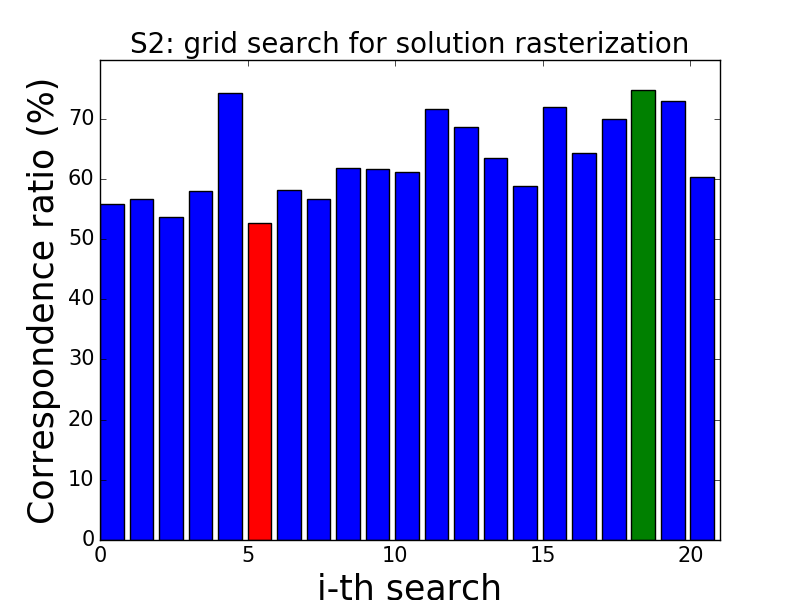} }
      \hbox{ \scriptsize \hspace{0.7cm} UAV 2: $\alpha_2=30, \sigma_2=30$}
      \hbox{ \scriptsize \hspace{1.5cm} $\alpha_2$-rise=20\% }
      \hbox{ \includegraphics[width=\linewidth]{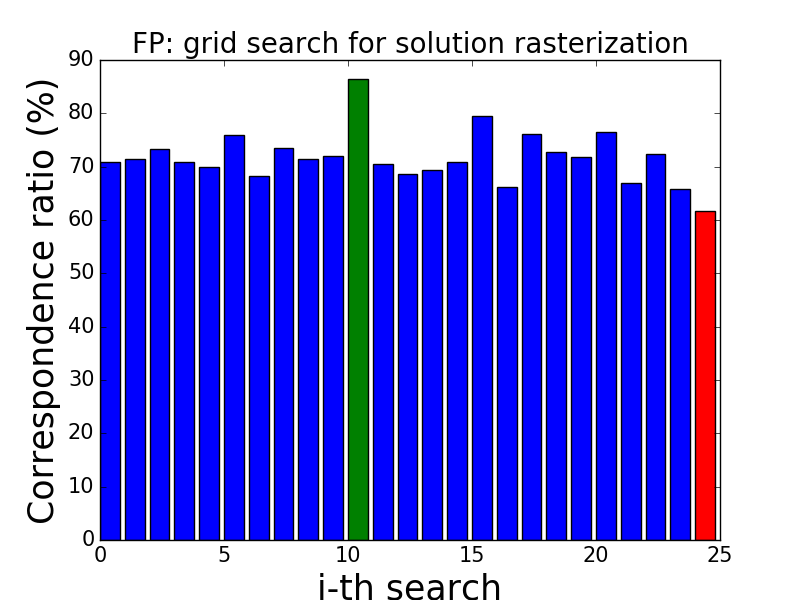} }
      \hbox{ \scriptsize \hspace{0.7cm} FP: $\alpha_2=40, \sigma_2=20$}
      \hbox{ \scriptsize \hspace{1.5cm} $\alpha_2$-rise=22\% }
    }
  }
  \caption{Correspondence ratio for several $\alpha_2$. Top: Wi-Fi UAV.
  Bottom: Monitoring UAV (AGP,FP). Setting $\alpha_2$ can boost the learning
  performance up to 29\%.}
\label{fig:grid_search}
\end{figure}

Figure \ref{fig:grid_search} shows the influence of the $\alpha_2$ Gaussian
parameter for the rasterisation of the expert solutions $\zeta_a$. When
$\alpha_2$ is high, $U$ is penalized for generating utility functions with
maxima $\hat{\zeta_a}$ even slightly different that the expert UAV locations
$\zeta_a$. This incites $U$ to strictly mimic the expert. However, when the
maxima $\hat{\zeta_a}$ are far from the expert's one, $U$ has no indication on
how to improve because all the learning signal is restricted to the
neighbourhood of $\zeta_a$. This can significantly slow down the training,
especially at the beginning of the optimization, and even prevent convergence.
As $\alpha_2$ decreases, the loss is more tolerant with $\hat{\zeta_a}$ that
differ from the exprt even though it can lead to lower correspondence ratio.
However, it allows $U$ to better recover from wrong utility functions. Finally,
there is an $\alpha_2$ value above which the rasterisation of the expert
placement can generate artificial maxima. 

We run a grid search on $(\alpha_2, \sigma_2) \in [20,60]^2$ with a step of 10
for both the Wi-Fi UAV and the monitoring UAV deployment problems.
We call $\alpha_2$-rise the difference between the correspondence ratio for the
worst (red) and the best (green) $\alpha_2$ parameter. Overall, we observe a
significant $\alpha_2$-rise of at least 20\%, and up to 29\% for the AGP. These
results reiterate the importance of data representation in learning from
demonstrations and shows that our loss definition ensures an efficient learning.
Empirically, we observe that setting $\alpha_2$ to the quarter of the average
distance between expert positions leads to satisfying results.

\section{Conclusion}

This paper shows how to leverage network experts previous efforts to automate
UAV deployment following the same expert policy. We learn a utility function
generator that autonomously learns relevant features for the expert when
solving the UAV placement problem. It aims at generating an approximation of
the expert's untold utility function which maxima are the optimal UAV
locations. To do so, it only relies on expert demonstrations made of UAV
previous missions and the expert solution. We model this generator with a CNN
for its representation space properties: the CNN can autonomously learn complex
and high-dimensional features relevant to the learning task which alleviate the
need to hand-craft them. This property allows the generation of utility
function that can solve NP-hard problems such as the Art
Gallery Problem. 

This learning method is tested on two types of UAV applications: dynamic Wi-Fi
networks and monitoring networks. Simulation experiments show that this method
can recover the expert's utility function without the need for the task model
or the closed-form expression of the utility. In both applications, the UAV
locations computed from the learned utility function match at least 75\% of the
expert's demonstrations. When they do not, they are still valid solutions to
complete the mission. This shows that the learned utility function generator
not only mimics the network expert but it also embeds the expert's untold
utility to provide alternative valid solutions.

\addtolength{\textheight}{-12cm}   %

\bibliography{root}

\begin{thebibliography}{10}

\bibitem{abadi2016tensorflow}
{\sc Abadi, M., Barham, P., Chen, J., Chen, Z., Davis, A., Dean, J., Devin, M.,
  Ghemawat, S., Irving, G., Isard, M., et~al.}
\newblock Tensorflow: a system for large-scale machine learning.
\newblock In {\em OSDI\/} (2016), vol.~16, pp.~265--283.

\bibitem{argall2009survey}
{\sc Argall, B.~D., Chernova, S., Veloso, M., and Browning, B.}
\newblock A survey of robot learning from demonstration.
\newblock {\em Robotics and autonomous systems 57}, 5 (2009), 469--483.

\bibitem{badrinarayanan2017segnet}
{\sc Badrinarayanan, V., Kendall, A., and Cipolla, R.}
\newblock Segnet: A deep convolutional encoder-decoder architecture for image
  segmentation.
\newblock {\em IEEE transactions on pattern analysis and machine intelligence
  39}, 12 (2017), 2481--2495.

\bibitem{borji2013state}
{\sc Borji, A., and Itti, L.}
\newblock State-of-the-art in visual attention modeling.
\newblock {\em IEEE transactions on pattern analysis and machine intelligence
  35}, 1 (2013), 185--207.

\bibitem{cgal:dhhk-gog-17a}
{\sc de~Castro, P. M.~M., Devillers, O., Hert, S., Hoffmann, M., Kettner, L.,
  Sch{\"o}nherr, S., Tifrea, A., and Gimeno, M.}
\newblock Geometric object generators.
\newblock In {\em {CGAL} User and Reference Manual}, {4.10}~ed. {CGAL Editorial
  Board}, 2017.

\bibitem{dumoulin2016guide}
{\sc Dumoulin, V., and Visin, F.}
\newblock A guide to convolution arithmetic for deep learning.
\newblock {\em arXiv preprint arXiv:1603.07285\/} (2016).

\bibitem{gonzalez2001randomized}
{\sc Gonz{\'a}lez-Banos, H.}
\newblock A randomized art-gallery algorithm for sensor placement.
\newblock In {\em Proceedings of the seventeenth annual symposium on
  Computational geometry\/} (2001), ACM, pp.~232--240.

\bibitem{guillen2016wifi}
{\sc Guillen-Perez, A., Sanchez-Iborra, R., Cano, M.-D., Sanchez-Aarnoutse,
  J.~C., and Garcia-Haro, J.}
\newblock Wifi networks on drones.
\newblock In {\em ITU Kaleidoscope: ICTs for a Sustainable World (ITU WT),
  2016\/} (2016), IEEE, pp.~1--8.

\bibitem{hayat2016survey}
{\sc Hayat, S., Yanmaz, E., and Muzaffar, R.}
\newblock Survey on unmanned aerial vehicle networks for civil applications: A
  communications viewpoint.
\newblock {\em IEEE Communications Surveys \& Tutorials 18}, 4 (2016),
  2624--2661.

\bibitem{kingma2014adam}
{\sc Kingma, D., and Ba, J.}
\newblock Adam: A method for stochastic optimization.
\newblock {\em arXiv preprint arXiv:1412.6980\/} (2014).

\bibitem{krause2008near}
{\sc Krause, A., Singh, A., and Guestrin, C.}
\newblock Near-optimal sensor placements in gaussian processes: Theory,
  efficient algorithms and empirical studies.
\newblock {\em Journal of Machine Learning Research 9}, Feb (2008), 235--284.

\bibitem{o1987art}
{\sc O'rourke, J.}
\newblock {\em Art gallery theorems and algorithms}, vol.~57.
\newblock Oxford University Press Oxford, 1987.

\bibitem{shewry1987maximum}
{\sc Shewry, M.~C., and Wynn, H.~P.}
\newblock Maximum entropy sampling.
\newblock {\em Journal of applied statistics 14}, 2 (1987), 165--170.

\bibitem{simonyan2014very}
{\sc Simonyan, K., and Zisserman, A.}
\newblock Very deep convolutional networks for large-scale image recognition.
\newblock {\em arXiv preprint arXiv:1409.1556\/} (2014).

\bibitem{tozoni2013practical}
{\sc Tozoni, D.~C., de~Rezende, P.~J., and de~Souza, C.~C.}
\newblock A practical iterative algorithm for the art gallery problem using
  integer linear programming.
\newblock {\em Optimization Online\/} (2013).

\bibitem{zeiler2011adaptive}
{\sc Zeiler, M.~D., Taylor, G.~W., and Fergus, R.}
\newblock Adaptive deconvolutional networks for mid and high level feature
  learning.
\newblock In {\em Computer Vision (ICCV), 2011 IEEE International Conference
  on\/} (2011), IEEE, pp.~2018--2025.

\end{thebibliography}
\bibliographystyle{acm}

\end{document}